\documentclass[superscriptaddress,prd,aps,twocolumn,floats,nofootinbib,amsfonts,amssymb,preprintnumbers,showpacs]{revtex4}
\usepackage{mathrsfs,graphicx,bm,amsmath}
\begin{document}

\title{\bf Nucleosynthesis and the variation of fundamental couplings}
\author{Christian M. M{\"{u}}ller}
\affiliation{Institut f{\"{u}}r Theoretische Physik, Philosophenweg 16, 69120 Heidelberg, Germany}
\author{Gregor Sch{\"{a}}fer}
\affiliation{Institut f{\"{u}}r Theoretische Physik, Philosophenweg 16, 69120 Heidelberg, Germany}
\author{Christof Wetterich}
\affiliation{Institut f{\"{u}}r Theoretische Physik, Philosophenweg 16, 69120 Heidelberg, Germany}
\date{May 19th 2004}

\newcommand{\be}{\begin{equation}}
  \newcommand{\ee}{\end{equation}}
\newcommand{\ba}{\begin{eqnarray}}
  \newcommand{\ea}{\end{eqnarray}}

\newcommand{\yp}{Y_{\textrm{p}}}
\newcommand{\yn}{Y_{\textrm{n}}}
\newcommand{\yhe}{Y_{\textrm{He}}}
\newcommand{\yd}{Y_{\textrm{d}}}
\newcommand{\yt}{Y_{\textrm{t}}}
\newcommand{\yht}{Y_{3}}

\newcommand{\li}{^7\textrm{Li}}
\newcommand{\he}{^4\textrm{He}}
\newcommand{\het}{^3\textrm{He}}

\newcommand{\aem}{\alpha_{em}}
\newcommand{\m}{M_{\bar{P}}}

\begin{abstract}
We determine the influence of a variation of the fundamental ``constants'' on the predicted helium abundance 
in Big Bang Nucleosynthesis. The 
analytic estimate is performed in two parts: the first step determines the dependence of the 
helium abundance on the nuclear physics parameters, while the second step relates those parameters to the fundamental 
couplings of particle 
physics. This procedure can incorporate in a flexible way the time variation of several couplings within a grand unified
 theory while keeping the 
nuclear physics computation separate from any model-dependent assumptions.

\end{abstract}

\preprint{HD--THEP-04-23}

\maketitle


\section{Introduction}

The possible time variation of the fundamental couplings touches a basic cornerstone of our understanding of particle physics 
\cite{Dirac:1937aa,Jordan:1939aa}.
It is a characteristic feature for cosmological models of quintessence -- combining the dependence of couplings on the value of the 
cosmological scalar field with the time variation of this ``cosmon'' field induces a time variation of the couplings with unfortunately 
unknown strength \cite{Wetterich:fk,Dvali:2001dd,Wetterich:2002ic}.

Recent observations of quasar absorption spectra by Webb et. al.  \cite{Webb:1998cq,Murphy:2001nu,Murphy:2003hw}
 have suggested that the electromagnetic fine structure ``constant''
might vary over cosmological timescales, $\Delta \aem/ \aem = -0.54(12) \times 10^{-5}$ for $z \approx 2$. However, 
other groups did exclude 
such a variation with high statistical significance \cite{Srianand:2004mq,Chand:2004ct,Quast:2003qu,Bahcall:2003rh}. 
Also systematic effects, such as the evolution of isotope ratios \cite{Ashenfelter:2003fn,Ashenfelter:2004kk} could have an impact on 
these measurements. While the reality of a  variation of $\aem$ in QSO absorption lines is still in dispute we need to gain an overview 
of other possible effects of a variation of the fundamental couplings on cosmological observations.
 
A central element of modern cosmology is Big Bang Nucleosynthesis (BBN). Actually, the bounds on the variation of  $\aem$ 
at $z\sim 3$ do not say much about the possible size of a variation $\Delta \aem/ \aem $ at the time of BBN, 
around $z \sim 10^{10}$. Furthermore, a major issue is the complex interplay of the variation of several couplings on 
the outcome of the element 
synthesis. While there are a number of recent investigations into the bounds of a   
variation of $\aem$  or other single parameters for BBN 
\cite{Ichikawa:2004ju,Yoo:2002vw,Copi:2003xd,Ichikawa:2002bt,Dmitriev:2003qq,Kneller:2003xf,Wetterich:2002ic}, 
we will follow a different 
approach which determines element abundances in a model independent way. For a review of current limits on fundamental
couplings see \cite{Uzan:2002vq}.

To determine light element abundances in the absence of time varying couplings, one needs to know the particle 
masses and the reaction rates 
of the relevant nuclear processes (from laboratory experiments). The time evolution of the scale factor $a(t)$ 
only depends on the number of 
relativistic particles or, more precisely, on the ratio $\rho/ T^{4}$ for energy density and temperature during BBN.
Excellent numerical codes for BBN abundance prediction exist \cite{Wagoner:1966pv,Wagoner:1972jh,Kawano:1988vh,Kawano:1992ua}. 
An essential 
parameter for these computations 
is the baryon to photon ratio $\eta$. Taking the value from WMAP measurements \cite{Spergel:2003cb}, 
 $\eta= 6.14 \pm 0.25 \times 10^{-10}$, yields a prediction of the helium abundance $\yhe= 0.2484^{+0.0004}_{-0.0005}$
 \cite{Cyburt:2003fe}. 
The observational 
determination varies among different groups. Izotov and Thuan \cite{Izotov:2003xn} quote two different values for two different 
equivalent width samples of spectra. The one is $\yhe= 0.2421 \pm 0.0021$ and the other $\yhe = 0.2444 \pm 0.0020$. If we were to 
calculate $\eta$ from those samples we would get $\eta=3.4 ^{+0.7}_{-0.6} \times 10^{-10}$ for the first and 
$\eta = 4.0^{+1.1}_{-0.5} \times 10^{-10}$ 
for the second quote. Another estimate for $\yhe$ was obtained by Fields and Olive \cite{Fields:1998gv}  
$\yhe = 0.238 \pm 0.002 \pm 0.005$, where the second is the systematic error (not quoted by Izotov and Thuan). 

These results show some tension between theory and observation. 
This discrepancy is likely due to systematic errors which are not fully understood. For instance the assumptions made 
about the extragalactic HII regions differ among several groups. 

Increasing the number of light species which are effective at BBN (e.g. more neutrinos) would enhance $\yhe$ and
 only worsen the discrepancy. 
(This also holds for the possible presence of early dark energy \cite{Wetterich:fm,Caldwell:2003vp}.) If a mechanism
 for \textit{decreasing} $\yhe$ has 
to be found 
the time variation of fundamental couplings seems to be a particularly plausible candidate \cite{Wetterich:2002ic}.
 Effects of the variation of
 the weak and strong scales or some dimensionless coupling on BBN  have been discussed long ago 
\cite{Wetterich:fk,Campbell:1994bf,Kolb:1985sj}.
One may therefore try to estimate the allowed variation of couplings at a very early time in cosmology.

Confidence limits on the variation of couplings or parameters in the framework of BBN always assume an underlying
model. However, the confidence regions determined from a model where only $\aem$ varies are meaningless if 
one wants to employ a model where other couplings, such as the weak scale, are allowed to change. 
In a Grand Unified Theory (GUT) framework, not only does the electromagnetic interaction vary, but also weak and strong interactions. 
The details of how these are connected depend on the specific GUT and the variation of the unified couplings and mass scales of
 spontaneous 
symmetry breaking. 
The present BBN limits on time varying couplings are difficult to compare due to this strong implicit model dependence. It is therefore 
essential to formulate the BBN estimates in a way that is as model independent as possible. 
This should facilitate the comparison between 
different assumptions on the time variation of fundamental couplings.

Our concept to solve this problem  is as follows. 
First we describe our general assumption, namely that the deviations are small, and explain how we linearize the problem 
(Section \ref{section_lin}). 
Rather than using a numerical computation we will use an analytic approximation 
to determine the variation of $\yhe$ -- this will help us to better understand how $\yhe$ depends on fundamental couplings.
 We emphasize that relative errors for the \textit{relative variation} below 10 \% are acceptable in contrast to the 
much higher required precision for the total abundance. 
In the first step of our analysis we estimate (Section \ref{section_he}) how the results of a BBN calculation depend 
on seven parameters $X_{i}$ characterizing nuclear physics (Eq. (\ref{param_phys})).
 
In a second step (Section \ref{section_fund}) we determine the dependence of the parameters $X_{i}$ on the relevant dimensionless 
``fundamental'' particle physics parameters $G_{k}$ (Eq. (\ref{param_fund})), which characterize the standard
 model of electroweak, strong and 
gravitational interactions. The connection between both is formulated in form of a ``transfer matrix'' $f_{ik}$. The advantage of this 
separation is the possibility to compute $f_{ik}$ without invoking BBN whereas the first step does not use any assumption 
about the particle physics -- nuclear physics connection. The two issues can therefore be dealt with independently. 
Any improvement on the 
estimate of the dependencies in the first step can be propagated to the fundamental couplings without repeating the whole calculation. 
In the same spirit 
one may discuss a third step which relates the standard model parameters to the GUT parameters 
\cite{Ellis:1989as,Calmet:2001nu,Calmet:2002ja,Langacker:2001td,Wetterich:2002ic}. 
Typically, this induces relations between the relative variations of the standard model parameters $\Delta G_k/G_k$.
As an example, we estimate in Section \ref{section_example} the variation $\Delta \yhe/ \yhe$ for two models for the connection between 
the $\Delta G_k/G_k$ within GUT models.
The size of the effect depends strongly on these models. Keeping the ratio between the characteristic scales for the weak and 
strong interactions fixed we obtain $\Delta \yhe/ \yhe= 35.0 \; \Delta \aem/\aem$ whereas for a fixed ratio between
 the weak scale and the 
GUT scale one finds  $\Delta \yhe/ \yhe= 130 \;\Delta \aem/\aem$.


\section{Linearization} \label{section_lin}
The success of BBN motivates the basic assumption of this paper, namely that the relative time variation of the fundamental 
constants between nucleosynthesis and the present epoch is small. We can then linearize in the relative variation of the fundamental 
parameters $\Delta G_{k}/G_{k}$ and use for $G_k$ the values extracted from laboratory experiments. We express the relative
 change of the 
helium abundance as
\be  \label{eq_sumgk}
 \frac{\Delta \yhe}{\yhe}=\frac{\yhe(G+ \Delta G)- \yhe}{\yhe} =\sum_k c^{(G)}_k \frac{\Delta G_k}{G_k}.
\ee
Here $\yhe$ corresponds to the helium abundance computed in absence of a cosmological time variation of couplings, assuming that only 
standard model particles (with three neutrinos) contribute to the energy density at BBN. 
Our aim is to determine the coefficients $c^{(G)}_{k}$ which relate the fundamental parameters to the change in $\yhe$.

In this paper we will consider the effects of the variation of six dimensionless quantities 
\ba \label{param_fund}
G_{k}=( && \m/ \Lambda_{QCD},\;\, \aem, \;\, \langle \phi \rangle / \Lambda_{QCD},  \nonumber \\
 && m_e / \Lambda_{QCD}, \;\,  m_q / \Lambda_{QCD}, \;\, \Delta m / \Lambda_{QCD}).
\ea
Here $\Lambda_{QCD}$ is the characteristic mass scale of the strong interactions which dominates the mass of the nucleons and 
the strong interaction rates whereas $\langle \phi \rangle$ is the Fermi scale (vacuum expectation value of the Higgs
 field) relevant for the 
weak interactions. The strength of the gravitational interactions is given by the (reduced) Planck mass $\m$ and $m_e$ is the
 electron mass. 
The up- and down-quark masses $m_u$, $m_d$ are reflected in $m_q= (m_u +m_d)/2$ and $\Delta m = m_d - m_u$.
In combination with $\langle \phi \rangle/\Lambda_{QCD}$ the three last mass ratios could be replaced by the relevant Yukawa couplings 
$h_e$,$h_u$,$h_d$. We emphasize that only ratios of mass scales are observable and have cosmological significance 
\cite{Wetterich:fk,Wetterich:fm,Wetterich:2002ic}.

For a given model for the time variation of the fundamental constants the variations $\Delta G_{k}/G_{k}$ are typically related to each 
other. For example, 
we may assume a unified theory (GUT) and vary only the gauge coupling at the unification scale $M_{GUT}$, while keeping $G_{3,4,5,6}$
fixed. This results in \cite{Wetterich:2002ic}($\Delta G_{3,4,5,6}=0$)
\be \label{eq_mp_aem}
\frac{\Delta(\m / \Lambda_{QCD} )}{\m / \Lambda_{QCD}} =  - \frac{\pi}{11} \frac{\Delta \aem}{\aem^2}= 
- \frac{\pi}{11} \frac{\aem^{BBN}-\aem}{\aem^2}.
\ee
Then only a single independent varying coulping is left that we may choose as $\Delta \aem/\aem$.

For practical reasons we  will work in a frame where we keep the strong scale 
$\Lambda_{QCD}$ fixed. This can be achieved by an appropriate Weyl scaling \cite{Wetterich:fk,Wetterich:fm} and will result in a time 
dependence of the reduced Planck mass $\m$. This particular frame  can be understood as a rescaling of the cosmological ``clock'' $\m$ 
which compensates for the constant strong interactions. In a frame with fixed $\m$ the strong interaction scale  $\Lambda_{QCD}$
would vary with time.

As mentioned before our computation of the coefficients $c^{(G)}_{k}$ proceeds in two steps. We first consider the 
dependence of $\yhe$ on the characteristic quantities for nuclear decays and reactions, also referred to as ``nuclear physics
 parameters'' 
\be \label{param_phys}
X_i= (\m,\; \aem, \; \langle \phi \rangle, \; m_e, \; \tau_n, \; Q, \; B_d),
\ee
according to  
\be \label{coeff_eq}
\frac{\Delta \yhe}{\yhe}=\sum_i c^{(X)}_i \frac{\Delta X_i}{X_i}.
\ee
Here, $\tau_n$ is the neutron lifetime, $Q$ the neutron proton mass difference and $B_d$ the deuteron binding energy.
 (We keep $\Lambda_{QCD}$ 
fixed -- otherwise the dimensionful parameters have to be multiplied by appropriate powers of $\Lambda_{QCD}$.) We emphasize 
that at this stage the effect of the variation of, say, $\aem$ is computed at fixed values of $X_{1,3,4,5,6,7}$. The computation 
of the coefficients $c^{(X)}_{i}$ involves the details of nuclear 
physics, i.e. reaction rates etc. .

A second step translates $\Delta X_i / X_i$ into the variation of the fundamental couplings 
\be
 \frac{\Delta X_i}{X_i} = \sum_k f_{ik}  \frac{\Delta G_k}{G_k},
\ee
with
\be  \label{eq_matrix}
f_{ik} = \frac{\partial \; ln X_i}{\partial \; ln G_k}.
\ee
This step involves the connection between nuclear physics and particle physics, namely the dependence of $\tau_n$, $Q$ and $B_d$ 
on the couplings $G_{2,3,4,5,6}$. (Obviously, one has $f_{ik}= \delta_{ik}$ for $i=1...4$ and $k= 1...6$.)
For known $f_{ik}$ the coefficients $c^{(G)}_{k}$ follow from $c^{(X)}_{i}$ as 
\be \label{eq_ck}
c^{(G)}_k = \sum_i c^{(X)}_i f_{ik}.
\ee

Before proceeding to estimates of the various coefficients we list our results for the dependence of the helium abundance on the nuclear 
physics parameters (Section \ref{section_he}) in Table \ref{coeff_phys}.

\begin{table}[h] \caption{\label{coeff_phys} Coefficients $c^{(X)}_i$ for nuclear physics parameters}
  \begin{ruledtabular}
    \begin{tabular}{cccccccc}
      variable &  $\m$ & $\aem$  & $\langle\phi\rangle$  & $m_e$ & $\tau_n$  & $Q$  & $B_D$ \\ \hline            
      coeff. & -0.81 & -0.043 & 2.4  & 0.024 & 0.24 & -1.8 & 0.53   \\
    \end{tabular}
  \end{ruledtabular}
\end{table}
\noindent 
The dependence on the particle physics parameters is shown in Table \ref{coeff_fund}. This Table allows for a quick inspection
 of the impact of 
various parameter variations. We observe a particularly strong influence of a possible variation of $\langle \phi \rangle$ and $\Delta m$
\cite{Wetterich:2002ic}.


\begin{table}[h] \caption{\label{coeff_fund} Coefficients $c_k^{(G)}$ for fundamental couplings}
  \begin{ruledtabular}
    \begin{tabular}{cccccccc}
      variable &  $\m$ & $\aem$  & $\langle\phi\rangle$ & $m_e$ & $m_q$  & $\Delta m$ \\ \hline            
      coeff. & -0.81 & 1.94 & 3.36  & 0.389 & -1.59 & -5.358   \\
    \end{tabular}
  \end{ruledtabular}
\end{table}


\section{Helium abundance}   \label{section_he}

In this section we want to derive a semi-analytic estimate for the primordial helium abundance $\yhe$. In doing so we will 
follow the approach of Esmailzadeh, Starkman and Dimopoulos \cite{Esmailzadeh:1990hf}, hereafter ESD. They estimate the BBN abundances
via quasi static equilibrium and fixed point conditions. This approach should be sufficient for an understanding of the effect of small 
variations. Of course, the estimate of the coefficients $c^{(X)}_i$as well as the determination of the corresponding errors 
would benefit 
from a systematic numerical investigation using the BBN codes.

In the primordial universe, at temperatures above several MeV, the abundances of protons and neutrons are in thermal 
equilibrium. Protons are converted into neutrons and vice versa. The latter process has a reaction rate given by \cite{Weinberg}
\ba \label{eq_gamma}
\Gamma_{n \rightarrow p}&=& A \int dx \;\, x^2 \left(1- \frac{m_e^2}{(Q+x)^2} \right)^{\frac{1}{2}} (Q+x)^2  \nonumber \\
&&(1 + e^{(x/T)})^{-1} (1+ e^{-(Q+x)/T})^{-1}. 
\ea
The integral runs from $-\infty$ to $+\infty$ with an energy gap between $-Q-m_e$ and $-Q+m_e$ with
$Q$ being the proton neutron mass difference. Here $A \sim \langle \phi \rangle^{-4}$ is the 4 point transition probability
 in Fermi-theory 
which depends on the axial and vector couplings $c_V$ and $c_A$. For simplicity we will work with constant $c_V$ and $c_A$.

We now assume that this reaction freezes out at a temperature $T^*_n$ when the Hubble expansion is comparable to 
this reaction rate, i.e.
\be \label{eq_freezeout}
\Gamma_{n \rightarrow p}(T^*_n) = b \; H(T^*_n). 
\ee
Note that an equally well justified assumption would be to include the reaction rate $\Gamma_{p \rightarrow n}$ in this condition. 
The insufficient information of our simple approach (as compared to a more complete treatment by  a solution of the Boltzmann equation) 
is accounted for by the unknown factor $b$. At the end we fix $b$  so that we obtain the same $\yhe$ as 
predicted by full numerical codes \footnote{For $b=1$ we obtain a $\he$ abundance that deviates by about 10 percent from the value 
$\yhe =0.2484$ found with a fully  numerical computation using the WMAP value for $\eta$ \cite{Cyburt:2003fe}. In order for our analytic 
approximation to yield the $\yhe$ predicted numerically  we use $b=1.22$.}.  
The Hubble parameter $H$ is given by the Friedman equation for a radiation dominated universe \footnote{We do not consider changes
 in the 
expansion rate due to changes in baryon or electron masses.}
\be \label{eq_fried}
H^2= \frac{ \rho}{3 \m^2}, \;\;\;\;\;\; \rho= g_* \ \frac{\pi^2}{30} T^4,
\ee
with an effective number of degrees of freedom $g_*= 10.75$ before positron-electron annihilation.
The freeze out temperature of the neutrons, with no change in fundamental couplings, obtains as $T^*_n = 0.77 \ \textrm{Mev}$ and  
the neutron concentration at freeze out can then be calculated as
\be
\yn^* = \frac{1}{1+ e^{Q/T^*_{n}}}= 0.158 \; .
\ee

Following the ``freeze out'' of the neutron to proton ratio the neutrons decay, thereby further changing $Y_n$ for $T < T^*_n$. 
After a short time the synthesis of 
deuterium and tritium starts which subsequently leads to the production of helium. Since almost all existing neutrons
end up in helium, we need to know how many neutrons remained when helium was synthesized in appreciable amounts.
We will assume that the neutrons decay freely until a time $t_f$ when helium formation starts to dominate over the neutron decay
 process, i.e. 
\be \label{eq_cond}
2 \dot{Y}_{\textrm{He}}(t_f)= - \dot{Y}_{\textrm{n}}(t_f).
\ee
The final $\he$ abundance is then estimated by
\be
\yhe = \frac{1}{2} Y_n(t_f) = \frac{1}{2} Y_n^* e^{-(t_f/\tau_n)}.
\ee
It depends on the couplings via $Q$, $T^*_n$, $\tau_n$ and $t_f$. In turn, $T^*_n$ depends 
on $A \sim \langle \phi \rangle^{-4}$, $Q$, $m_e$ 
and $\m$ via Eqs. (\ref{eq_gamma}), (\ref{eq_freezeout}) and (\ref{eq_fried}).

We need an estimate of $t_f$. The by far dominant process for helium production is the reaction \cite{Smith:1992yy} 
\be 
d\,+\,t \rightarrow  \, \he  +n \; . 
\ee             
To write down the equations governing the abundances comprised of several reactions we will adopt the 
notation of ESD who abbreviate a reaction rate 
\be
\alpha + \beta \rightarrow \gamma + \delta
\ee
as $[\alpha \beta \gamma \delta]$.
The condition for the time until which the neutrons decay is given by
\be \label{he}
2 \yd \yt [dtn\alpha] = \frac{1}{\tau_n} \yn^* e^{-t_f/\tau_{n}}.
\ee
To compute the time when this relation is satisfied we need to know the 
abundance of deuterium($\yd$) and tritium($\yt$) as well as the reaction rate $[dtn\alpha]$. In the temperature range we are
 considering, 
deuterium can be assumed to be in thermal equilibrium and hence its abundance is given by the Saha equation \cite{Kolb:vq}
\be  \label{eq_deuterium}
\yd = 8.15  \, \left(\frac{T}{m_n}\right)^{3/2}\eta \ e^{B_d/T}  \; Y_n Y_p ,
\ee
with the proton abundance being $\textrm{Y}_{p} \approx (1-\yn) $, $m_n$ the neutron mass.  

The estimate of $\yt$ is more involved and also requires knowledge of the abundance $\yht$ for $^{3}\textrm{He}$.
The tritium concentration is established by the reactions 
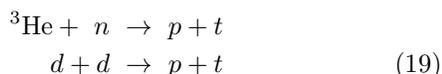
\ba
 ^3\textrm{He} + \ n &\rightarrow&  p + t  \nonumber \\
d+d &\rightarrow& p + t   
\ea
creating and 
\be
t+d \rightarrow  \, ^4\textrm{He} +n  
\ee
annihilating tritium. Other reactions are subdominant by at least 2 orders of magnitude (as can be verified from  \cite{Smith:1992yy}) 
and are therefore neglected. Close to thermal equilibrium the fixed point condition \cite{Esmailzadeh:1990hf} leads us
 to an equation for $\yt$ 
of the form:
\be  \label{eq_tritium}
\yt = \frac{\yn \,\yht \ [n3pt] + \yd \ \yd \ [ddpt]} 
{\yd \ [dtn\alpha] } .
\ee 
Likewise, we can write down the dominant processes for the $\het$ abundance. Invoking 
the fixed point condition yields
\be  \label{eq_he3}
\yht = \frac{ \yd \ \yp [pd3\gamma] + \yd \ \yd [ddn3]}
{\yd \  [d3p\alpha] + \yn \ [n3pt] }.
\ee
From Eqs. (\ref{eq_deuterium}), (\ref{eq_tritium}) and (\ref{eq_he3}) we can determine the abundance of deuterium, tritium 
and helium-3 as a
function of $T$ and $Y_n$. In turn, temperature and time are related by the background cosmology and $Y_n= Y^*_n \; e^{-t/\tau_n}$. Eq. 
(\ref{eq_cond}) now determines $t_f$. 
  
The dependence of $\yhe$ on the various parameters cannot be solved analytically. 
In the linear approximation, however, the computation of the response coefficients $c^{(X)}$ is straightforward. For this purpose
 we assume 
that all strong interaction rates are determined by the strong interaction scale $\Lambda_{QCD}$. At this point we 
benefit from our particular 
frame with constant $\Lambda_{QCD}$ which implies that we can use constant rates $[dtn\alpha]$ etc., except for small 
electromagnetic effects.

The results of this computation can be found in Table \ref{coeff_phys}. They are plausible in the sense that they resemble 
what one would 
expect from simple arguments.
Increasing the Planck mass gives a slower expansion rate, resulting in a later freeze out of weak interactions, hence less
neutrons are available for helium production. 
 Increasing the decay  time $\tau_n$ of the free neutrons leaves more neutrons to be converted 
into helium since effectively all neutrons end up 
being bound in helium. Increasing $\langle \phi \rangle$ results in a decrease of the Fermi interaction $G_F$, hence weak
interactions freeze out earlier resulting in an increase in $\yhe$. 
Changing $Q$ results in a different neutron-proton ratio at 
freeze out and also in modified weak rates due to changes in the available phase space. If we exclude the changes in 
phase space volume, the coefficient is $-1.4$ instead of $-1.8$. Thus, helium abundance is a decreasing function of the proton-neutron 
mass difference $Q$ as anticipated. 
Increasing the binding energy of the deuteron, $B_D$, results in earlier formation of helium and reduces the amount of 
neutrons decaying into protons.The influence of the electron mass is only through the phase space volume in the weak rates 
which is a very small effect for our purposes.

Changes in $\aem$ affect the nuclear reaction rates with the main effects being variations in the Coulomb barrier
for charge-induced reactions, final-state interactions, radiative capture and mass differences. We use the procedure of 
Bergstr\"{o}m, Iguri and Rubinstein (BIR) \cite{Bergstrom:1999wm} for computing the impact of varying $\aem$ on all nuclear 
reaction rates used in our computation, including the improvements of Nollett and Lopez \cite{Nollett:2002da}.
We use the rates of the
NACRE compilation \cite{NACRE} where available, otherwise we use those of Smith, Kawano and Malaney (SKM) \cite{Smith:1992yy}. For the 
process $^3\textrm{He}(n,p)t$ we use the fit of Cyburt, Fields and Olive \cite{Cyburt:2001pp}.
Since the analytic NACRE rate fits have a different 
expansion in terms of $T$ we have fitted the rates to the SKM functional form for use of the BIR treatment as described 
in \cite{Nollett:2002da}. 

Except for electromagnetic effects we have not taken into account any other effect that may change the reaction rates.



\section{Fundamental couplings}  \label{section_fund}
In this section we describe 
the relation between the fundamental couplings $G_k$ and the nuclear physics parameters $X_i$. This relation was 
expressed in the form of a matrix equation (\ref{eq_matrix}). We will now discuss what effects we took into account by explaining 
each row of the matrix $f_{ik}$ (see Table \ref{table_matrix}). 
Each coefficient $f_{ik}$ describes the response of the ``nuclear physics parameter'' $X_i$ when one varies a single parameter
 $G_k$, while
keeping the other $G_{l \neq k }$ fixed. For $i=1...4$ the parameters appear both in the lists of $X_i$ and
 $G_k$ and $f_{ik}= \delta_{ik}$ 
by virtue
of our definition. Also $\tau_n$, $Q$ and $B_d$ do not depend on $\m$ implying $f_{1k}=\delta_{1k}$. The nontrivial
 coefficients $f_{ik}$ for 
$i=5,6,7$ account for the dependence of $\tau_n$, $Q$ and $B_d$ on $\aem$, $\langle \phi \rangle$, $m_e$, $m_q$ and $\Delta m$.
 
The nucleon masses and nuclear binding energies depend on the quark masses and $\aem$. The dependence of the  neutron-proton
 mass difference 
on the fundamental couplings is given by (see \cite{Gasser:1982ap}): 
\be
Q= \left[ -0.76 \left(1+\frac{\Delta \aem}{\aem}\right)
 + 2.05 \left(1+ \frac{\Delta (\Delta m)}{\Delta m} \right) \right]\mbox{ MeV }.
\ee
From this we can determine $f_{62}$ and $f_{66}$.
Recent studies have suggested that the deuteron binding energy $B_d$ may 
increase with decreasing pion mass \cite{Epelbaum:2002gb,Beane:2002xf}. We may parametrize the dependence of $B_d$ on  $m_{\pi}$
 at fixed 
$\langle \phi \rangle$ by a linear fit \cite{Yoo:2002vw} and neglect the dependence on $\langle \phi \rangle$ at fixed $m_q$, $\Delta m$.
For the electromagnetic part we use the Monte Carlo simulation data of Pudliner et al. \cite{Pudliner:1997ck}. Hence 
the deuteron binding energy may be expressed in terms of the pion mass $m_{\pi} \propto  m_q^{1/2}$ and $\aem$ as
\be
B_d=B_{d}^0\left[ (r+1)-r \frac{m_{\pi}}{m_{\pi}^{0}} \right]
-0.018\frac{\Delta \aem}{\aem}\mbox{ MeV },
\ee
where $r$ is a parameter that varies between 6 and 10 and $B_d^0=2.225$ MeV is the deuteron binding energy as measured 
in the laboratory today. 

The neutron lifetime  is changed due to variations in the weak scale $\tau_n \propto G_{F}^{-2} \propto \langle \phi \rangle^4 $. 
Furthermore, a change in the phase space volume $f$ of free neutron decay
\be
f = \int^Q_{m_e} dq \;\, q^2 (Q-q)^2 (1- \frac{m_e^2}{q^2})^{1/2},
\ee
results in a dependence of $\tau_n$ on $Q $ and the electron mass $m_e$. Because $Q$ also depends on $\aem$ and $\Delta m$,
 $\Delta \tau_n$ 
will also have contributions from the variation of those parameters. A linear analysis then yields the corresponding
 entries for $\tau_n$: 
\be
\frac{\Delta \tau_n}{\tau_n} = 3.86 \frac{\Delta \aem}{\aem} +  4 \frac{\Delta \langle \phi \rangle }{\langle \phi \rangle}  
+ 1.52\frac{\Delta m_e}{m_e} - 10.4 \frac{\Delta (\Delta m)}{\Delta m}.
\ee

The entries of the matrix $f_{ik}$ can be found in Table \ref{table_matrix}. We have quoted the coefficients for the effects we discussed
above. When there is no contribution at all a zero is written. For some relations between the $G_k$ and the $X_i$ small effects 
are present but with negligible coefficients. To distinguish those from the others, we have left the matrix entry empty.
Having determined the transfer matrix we can calculate the dependence of $\Delta \yhe/\yhe$ on the fundamental parameters 
(see Eq. (\ref{eq_ck})). The results are shown in Table \ref{coeff_fund} above.


\begin{table}[h] \caption{\label{table_matrix} The matrix entries $f_{ik}$, corresponding to the coefficients relating relative changes 
in $G_k$ to relative changes in $X_i$.}
  \begin{ruledtabular}
    \begin{tabular}{ccccccc}
      parameter \footnote{The parameters are dimensionless, but we omitted the scaling by $\Lambda_{QCD}$} & $\m $ &  $\aem$ & 
$\langle \phi \rangle$ &  $m_e $ & $m_q $& 
     $\Delta m $ 
     \\ \hline            
     $\m$   &  1&0&0&0&0&0 \\ 
     $\aem$ & 0&1&0&0&0&0 \\ 
     $ \langle \phi \rangle$ & $ 0$&0&$1$&0&0&0 \\ 
     $m_e$ &  0&0&0&$1$&0&0 \\ 
     $\tau_n$ & 0&$3.86$&$4$&$1.52$&-&$-10.4$  \\
     $Q$ & $0$&$-0.59$&-&-&-&$1.59$  \\ 
     $B_d$ &  0&$-0.0081$&-&-& $-r/2$ &-  
    \end{tabular}
  \end{ruledtabular}
\end{table}


\section{Two GUT examples} \label{section_example}

In this section we present two examples based on GUTs for which we have expressed the changes in the fundamental 
parameters by the variation of
only one independent coupling and computed the resulting change in $\yhe$. 
The variation of the couplings is assumed to be due to a scalar field $\chi$ called the cosmon \cite{Wetterich:fk,Wetterich:fm}.
There are good arguments \cite{Wetterich:2002ic} that this field plays the role of quintessence \cite{Wetterich:fm} and its
 present potential 
and kinetic energy can be associated with the dark energy of the universe. For our considerations, however, we will not need 
any particular 
details of the evolution of the cosmon, except that its value at the time of nucleosynthesis was different from its present value. 

For the details of the derivation of how the fundamental constants change in a GUT scheme we refer the reader to 
\cite{Wetterich:2002ic,Wetterich:2003jt}.
Merely quoting the results, to one loop order the fundamental couplings as functions of the cosmon field $\chi$ are given by  
\cite{Wetterich:2002ic}
\ba \label{eq_as}
\alpha_{s}^{-1}(M_W) &=& \frac{4 \pi Z_{F}(\chi)}{\bar{g}^2}+ \frac{7}{2 \pi} \ln \zeta_{w}(\chi),  \\
\alpha_{w}^{-1}(M_W) &=& \frac{4 \pi Z_{F}(\chi)}{\bar{g}^2}+ \frac{5}{3 \pi} \ln \zeta_{w}(\chi),  \label{eq_aw}  \\
\alpha_{em}^{-1}(M_W) &=& \frac{32 \pi Z_{F}(\chi)}{3 \bar{g}^2}- \frac{5}{3 \pi} \ln \zeta_{w}(\chi),\label{eq_aem} 
\ea
where the W-Boson mass is $M_{W}(\chi)= \zeta_{w}(\chi) \; \chi$ and $Z_{F}(\chi)$ determines the renormalized grand 
unified gauge coupling 
($g^2_R= \bar{g}^2/Z_F$, $\bar{g}$ fixed). We normalize $\chi$ such that $M_{GUT}(\chi)= \chi$. In 
Eqs. (\ref{eq_as})-(\ref{eq_aem}) we can replace 
$M_W=g_w \langle \phi \rangle / \sqrt{2}$ by  $\langle \phi \rangle$. The relative variation of $g_w$ (or $\alpha_{w}$)
 induces only a correction 
of higher order in these relations.

As mentioned before, we will work in a frame in which the scale of the strong interaction is fixed such that the strong 
interaction rates are constant for our BBN estimate. We will consider two particularly simple scenarios where 
$\m(\chi)/ M_{GUT}(\chi)= \textrm{const.}$ with 
\be
\frac{\Delta \m / \Lambda_{QCD}}{\m / \Lambda_{QCD}} = - \Delta \ln \zeta_{w} + \Delta \ln (\langle \phi \rangle/ \Lambda_{QCD}).
\ee

 Furthermore, we also neglect 
the variation of the Yukawa couplings and hence the variations in $m_e$, $m_q$ and $\Delta m$ obey
\be \label{eq_me_phi}
 \frac{\Delta m_e}{m_e}=  \frac{\Delta (\Delta m)}{\Delta m}=  
\frac{\Delta m_q}{m_q}= \frac{\Delta \langle \phi \rangle }{\langle \phi \rangle}  .
\ee  
The effect of the variation of the cosmon field $\chi$ can now be expressed as a varation in the renormalized grand
 unified gauge coupling expressed by $Z_F$ and
a variation in $\ln \zeta_w$.

At this stage the two unknown quantities $\Delta \ln  Z_F$ and $\Delta \ln \zeta_w$ contain all relevant information about the 
unknown coupling of the cosmon to matter and radiation. For the present investigation we can simply use the relative variation
of the GUT-coupling $\Delta\ln Z_F$ and the ratio between weak and GUT scale $\Delta \ln \zeta_w$ as free parameters. 
For the running of $\aem$ at $\mu < M_W$ we have the relation
\be \label{eq_scales}
\aem (\mu)^{-1}=\aem (M_W)^{-1} + \frac{2}{3\pi} \sum_i Q_i^2 \; \ln \frac{M_W}{\mu},
\ee
where the $Q_i$ are the charges of the particles with masses in the range between $M_W$ and $\mu$. In our case this is given
by five quarks (top lies above $M_W$) in three colours plus 3 leptons, i.e.  $\sum_i Q^2_i=3 \times (8/9+ 3/9)+3 $.
Similarly, for the running of $\alpha_s$ below $M_W$ we include five quarks and associate $\Lambda_{QCD}$ with the scale
 where the one loop 
expression for $\alpha_s(\mu)$ diverges.

We can now express $\aem=\aem(m_e)$ and $\Lambda_{QCD}$ in terms of $\aem(M_W)$ and $\alpha_{s}(M_W)$. Thus they 
relate $\Lambda_{QCD}/ \chi$ and $\aem$ to $Z_F$ and $\ln \zeta_w$. The specific relation between the variations of
  $\Lambda_{QCD}/\m$ and $\aem$ 
depends on the variation of the weak scale $\ln \zeta_w$. Our two examples will either keep $\ln \zeta_w$ or  
$\ln (\langle \phi \rangle/ \Lambda_{QCD})$ fixed.

The first example is as simple as possible -- we also keep the weak scale fixed w.r.t. the strong scale. This will result in a 
$\chi$ independent ratio $\langle \phi \rangle /\Lambda_{QCD} \sim \textrm{const.}$ and leads to 
\be \label{eq_delaem_ex1}
\frac{\Delta \aem(M_W) }{\aem^2(M_W)}= -  \frac{88 \pi}{7} \frac{\Delta Z_F}{\bar{g}^{2}}
\ee
and
\be \label{eq_mp_aem_ex1}
\frac{\Delta \m/\Lambda_{QCD} }{\m/\Lambda_{QCD}}= - \frac{\pi}{11} \;\, \frac{\Delta \aem(M_W)}{\aem^2(M_W)}.
\ee

Since the weak scale is fixed we set $\Delta \langle \phi \rangle/ \langle \phi \rangle = 0$ in Eq.(\ref{eq_sumgk}).

In our first example one has $\Delta \aem^{-1}(m_e)= \Delta \aem^{-1} (M_W)$ which is related to the only unknown parameter
 $\Delta Z_F$ by Eq. 
(\ref{eq_delaem_ex1}). Eq. (\ref{eq_mp_aem_ex1}) results in 
\be \label{eq_ex1_number}
\frac{\Delta(\m / \Lambda_{QCD} )}{\m / \Lambda_{QCD}}=   -39.1 \;\, \frac{\Delta \aem}{\aem}.
\ee
The only nonvanishing entries in Eq. (\ref{eq_sumgk}) from $\Delta G_1$ and $\Delta G_2$ are therefore related and
 $\Delta \yhe/ \yhe$ can be 
expressed in terms of a single parameter that we may choose as $\Delta \aem /\aem$.  In order to get an idea of the 
 sensitivity we compute the 
value $\Delta \aem/\aem$ which would be needed in order to obtain a helium  abundance $\yhe =  0.24$  for $\eta$ 
corresponding to the central WMAP 
value. For our first example we find \footnote{ We have put $r=6$. The change in $\yhe$ is negligible if we choose $r=10$.}
\be \label{eq_result1}
\frac{\Delta \aem (m_e)}{\aem(m_e)}= -1.0 \;  \times 10^{-3}.
\ee 

In the second example we will set $\langle \phi \rangle / \chi = \textrm{const}$. For this setting we obtain
\be
\frac{\Delta \alpha_{em}(M_W)}{\aem^2(M_W)} = -  \frac{32 \pi }{3}  \frac{\Delta Z_F}{\bar{g}^{2}}
\ee
and
\be \label{eq_mpaem_ex2}
\frac{\Delta \m/\Lambda_{QCD} }{\m/\Lambda_{QCD}}=\frac{\Delta \langle \phi \rangle }{\langle \phi \rangle}= - \frac{\pi}{12} \; \,
\frac{\Delta \aem(M_W)}{\aem^2(M_W)}.
\ee  
 We also need to relate the different mass scales via Eq.(\ref{eq_scales}) 
giving
\be
\frac{\Delta \aem (m_e)}{\aem^2(m_e)}= \frac{\Delta \aem(M_W) }{\aem^2(M_W)}[1+ \frac{1}{18} \sum_i \tilde Q_i^2],
\ee
where $\sum_i \tilde Q_i^2 =2$ runs only over the three light quarks whose effect on the running of $\aem$ is cut off 
at $\mu \sim \Lambda_{QCD}$.

In Eq.(\ref{eq_sumgk}) we now have nonvanishing entries from $\Delta G_{3,4,5,6}$ as well, related by Eqs. (\ref{eq_me_phi}) and 
(\ref{eq_mpaem_ex2}) to $\Delta \aem / \aem$ giving
\be \label{eq_ex2_number}
\frac{\Delta(\m / \Lambda_{QCD} )}{\m / \Lambda_{QCD}}= 
\frac{\Delta \langle \phi \rangle}{\langle \phi \rangle}=  -32.3 \;\, \frac{\Delta \aem}{\aem}. 
\ee

The variation of $\Delta \aem $ which would be needed for $\yhe = 0.24$ is now reduced as 
compared to the first model,
\be \label{eq_result2}
\frac{\Delta \aem (m_e)}{\aem(m_e)}= -2.7 \; \times 10^{-4}.
\ee


\section{Discussion and Conclusion}

Big Bang nucleosynthesis offers an excellent testing ground for the time variation of the fundamental couplings, probing directly their 
values at a time close to the big bang or at high redshift.
This paper presents a general analysis how the primordial helium abundance depends on six ``fundamental couplings''
 as summarized in Table 
\ref{coeff_fund}. In grand unified (GUT) models the variations of the various couplings are interrelated. Typically, 
the dominant effect comes 
from a variation of the ratios $\m/ \Lambda_{QCD}$ or $\langle \phi \rangle/ \Lambda_{QCD}$ rather than from the direct influence of a 
variation of the fine structure constant \cite{Wetterich:2002ic}. This is easily seen by comparing Eqs. (\ref{eq_ex1_number}) and 
(\ref{eq_ex2_number}) with Table \ref{coeff_fund}. As noted before \cite{Wetterich:2002ic} the variation of the weak interaction scale 
$\langle \phi \rangle$ or the quark masses can play a very substantial role. We have checked that the impact of a variation 
of the strong and 
electromagnetic interaction rates is a rather minor effect. Discussing two different models gives us a certain handle to investigate the 
effect of cancellations -- compare Eqs. (\ref{eq_result1}) and (\ref{eq_result2}). Extending our analysis to the abundances 
of deuterium and 
lithium may be a way to (partially) lift the degeneracies between the variations of various couplings.

Excluding very particular cancellations we may infer from the approximate agreement between the WMAP-prediction and the 
observations of $\yhe$ 
a bound $|\Delta \aem/\aem (z= 10^{10})| < \textrm{a few times}\;  10^{-3}$. A typical size of a coupling variation that 
could explain the present
discrepancy between WMAP and the observed $\yhe$ would be in a range $\Delta \aem/\aem \approx (2-10) \times 10^{-4}$.
 
Obviously our treatment can be improved. The coefficients quoted in Table \ref{coeff_phys} could be estimated with higher 
accuracy by using
a full numerical code instead of our analytic estimate. Likewise, looking at Table \ref{table_matrix} one can see that 
there are some small
effects contributing to the matrix $f_{ik}$ which we have not included. 
Also, we have only investigated the change in the helium abundance. Stringent bounds on BBN are obtained from the primordial abundances 
of deuterium and lithium and it would be worth extending our analysis to these other light elements.


\begin{acknowledgements}
We gratefully thank J{\"{o}}rg J{\"{a}}ckel for helpful discussions. 
C.M. M{\"{u}}ller and G. Sch{\"{a}}fer are supported by 
GRK grant 216/3-02.
\end{acknowledgements}



\bibliographystyle{unsrt}

\end{document}